\documentclass[
aps,%
10pt,%
final,%
notitlepage,%
oneside,%
twocolumn,%
nofootinbib,%
superscriptaddress
]{revtex4-2}

\usepackage{psfrag}%
\usepackage{color}%
\usepackage{rotating}%
\usepackage{graphics}%
\usepackage{caption}
\usepackage{multirow}

\makeatletter
\renewcommand \thefigure{\@arabic\c@figure} % Arabic fig numbers.
\renewcommand \thetable{\@arabic\c@table} % Arabic table numbers.
\setcitestyle{authoryear,aysep={,},yysep={;},semicolon,round}

\makeatletter
\renewcommand \thefigure{\@arabic\c@figure} % Arabic fig numbers.
\renewcommand \thetable{\@arabic\c@table} % Arabic table numbers.
\usepackage{xurl} % Allows to bleak lonk links, must be before hyperref.
\usepackage[breaklinks]{hyperref}

\hypersetup{
    colorlinks=true,
    linkcolor=blue,
    filecolor=magenta,
    urlcolor=blue,
    citecolor=blue,
    breaklinks=true % This is to tell hyperref to break long links.
}

\begin{document}

%\selectlanguage{russian}%

%\noindent {\it Astronomy Reports, 2025, vol. 102, No. 12}
%\bigskip\bigskip  \hrule\smallskip\hrule
%\vspace{35mm}

\title{Period Change of the Binary System WR+O V444 Cyg: Updated Ephemeris Formula}

\author{\bf %\copyright $\:$  2025.
\quad \firstname{I.~I.}~\surname{Antokhin}}%
\email{igor@sai.msu.ru}
\affiliation{%
Lomonosov Moscow State University, Sternberg Astronomical Institute, Moscow, Russia}

\begin{abstract}

%\centerline{\footnotesize Received September 30, 2025;$\;$}
%\centerline{\footnotesize revised 13 November, 2025;$\;$}
%\centerline{\footnotesize accepted 18 November, 2025}
\bigskip

\mbox{V444~Cyg} is a \mbox{WN5+O6\,V} eclipsing binary system that exhibits a secular variation in its orbital period due to the loss of matter from the Wolf$-$Rayet star through its powerful stellar wind. This makes it possible to obtain a dynamical estimate of the WR~star mass-loss rate with minimal modeling assumptions. Numerous studies have been published on this topic. Unfortunately, over time, they have accumulated various flaws due to the authors' differing use of previously published light curves. In this paper, we have critically analyzed all published data, added new data obtained by us, and present a table containing all currently known times of the primary minimum, found in a uniform manner and based on independent original data. Using this table, we updated the value of the parameters of the quadratic formula describing the times of the primary minimum. The found rate of orbital period change is \mbox{$\dot{P} = 0.134\pm 0.003$~s/year}, and the corresponding value of the WR star mass-loss rate is \mbox{$\dot{M}_{\rm WR} = (6.82\pm 0.26) \times 10^{-6}\,M_\odot$/year}.

\medskip


\end{abstract}

\maketitle

\section{Introduction}

\mbox{V444~Cyg} (WN5+O6\,V, $V\simeq 8.0^m$, $P\simeq 4.2^d$)~ is one of the best-studied binary systems of the \mbox{WR+O} type. This is due to the facts that spectral lines of both components are visible in its spectrum, it is eclipsing, and the distance between the components is not very small \mbox{($\sim\!36\,R_\odot$)}, so that the light curves and spectra are not too burdened by the effects of proximity. At the same time, its orbital period is relatively short, which is convenient from the point of view of accumulating observational data. The period of the system increases with time, the main reason for this is the change of the orbital angular momentum due to the loss of matter through the stellar wind of the~WR~ component. The change of the period makes it possible to estimate an important parameter of the~WR~ wind --- the mass-loss rate. Alternative estimation methods include, for example, measuring the radio flux of a single WR star or analyzing its spectrum with a wind radiative transfer model (see, e.g.,~\citealp{nugis00, hamann19}). The radio flux of a WR star is determined by the free-free emission of the wind material, which depends on the square of its density. The shape of the emission spectral lines of WR stars (especially their red wings) also depends on the square of the wind density. Thus, both of these methods are critically dependent on the adopted wind inhomogeneity model, which is insufficiently studied and is usually described by very simple approximations, such as the so-called ``micro-inhomogeneities''. Consequently, both methods tend to significantly overestimate the mass-loss rates. On the other hand, as shown in~\cite{khal74}, estimating the mass loss rate from the change of the orbital period is independent of the detailed wind structure and is thus more reliable.

Dozens of papers, beginning in the 1960s, have been devoted to studying the orbital period variations of \mbox{V444~Cyg}. A critical review of these papers is provided in \mbox{section~\ref{sec:review}}. \mbox{Section~\ref{sec:ephem}} presents a table of the times of the primary minimum of the V444 Cyg light curve, obtained by carefully filtering data from previous authors and by using our own measurements. The section also presents updated parameters of the quadratic ephemeris formula, along with the rate of orbital period change, and the most probable mass-loss rate of the WR star. The conclusions of the paper are presented in section \mbox{\ref{sec:concl}}.

\section{Review of previous research}\label{sec:review}

The most reliable way to study the period variations of a binary system is to measure the times of minima of its light curve. In principle, one could also use radial velocity curves~\citep{khal84, under90, shap23} or, for example, variations in the Stokes parameters from polarization observations~\citep{louis93}; however, the accuracy of determining the times of minima from such data is significantly lower. Therefore, in this paper, we will limit ourselves to photometric data.

The binarity of \mbox{V444~Cyg} was first discovered by \cite{wilson39} from spectral observations (see the review of spectral observations in~\citealp{shap23}). In subsequent years, several papers appeared devoted to the photometric variability of the system. \cite{gaposh41} published the first photographic light curve in ``yellow'' and ``red'' bands, folded with the orbital period. The measurements were made using photographic plates from the Harvard University survey (the modern name of the plate collection is Harvard Plate Stacks). Unfortunately, the author did not publish the original data and did not indicate their mean epoch (only the observation date of the first plate is known~--- 1898, and the number of plates used~--- 1084). It is highly likely that, since the survey was conducted systematically over many years (until the end of the 1990s), these plates were obtained more or less uniformly over the interval from 1898 to the submission date of the Gaposhkin's article for publication (1940). This assumption is supported by the fact that the author determined 84~times of the primary minimum from these data (which would be meaningless if the observations covered a small time interval). Unfortunately, Gaposhkin does not give these times in his article, only stating that he did not detect a period change. Therefore, in a later paper by \cite{khal84}, the mean Gaposhkin light curve folded with the orbital period was used to calculate just one time of the primary minimum, to which the mean epoch of the interval \mbox{1898$-$1940} was assigned. In current paper, we also use this time of the primary minimum.

New broadband photoelectric light curves were obtained by \cite{kron43} in the ``blue'' band (\mbox{$\lambda_{\rm eff}\sim 4500$~\AA})~ and by \cite{kron50} in the ``red'' band (\mbox{$\lambda_{\rm eff}\sim 7200$~\AA}), by \cite{hiltner49} (\mbox{$\lambda_{\rm eff}\sim 3550$~\AA}), and also by \cite{cher65} ({$B$, $V$} bands) and \cite{gusein65} ($U$, $B$, $V$ bands). The orbital phase-folded light curves provided by these authors were used by \cite{sem68} in an attempt to determine whether the orbital period of the system changes. The author found the phase shift between the light curves and the corresponding times of minima using \cite{hertz1919} method in the version proposed by \cite{kwee56}. She came to the conclusion that ``the period of \mbox{V444~Cyg} does not decrease''. Having mentioned Gaposhkin's observations, she notes that they do not contradict this conclusion. In the listed papers, it was also shown that the times of minima of the light curve do not depend on the effective wavelength of the (optical) filter used.

\cite{cher_khal72, cher_khal73} were the first to point out that, since the emission lines of the WR star are formed in its extended wind and contribute significantly to its emission, determining the parameters of the system and its components from light curves obtained in broadband filters (e.g., $U$, $B$, $V$) requires modeling the ionization state in the wind. This problem was not solved at that time (and remains not completely solved at present days). Modeling the system is greatly simplified by using narrowband filters such that within their passbands spectral lines are (nearly) absent. The authors obtained such observations in several narrow bands (\mbox{$\lambda~4244$}, 4789, 5806, 6320, \mbox{7512~\AA}, with the FWHM from \mbox{50~\AA} to \mbox{109~\AA}, respectively). They showed that the observed shape of primary minima (the WR star is in front, the shape of the minimum is affected by the opacity of the WR wind) in different bands is consistent with the hypothesis that the main source of the WR wind opacity in the optical continuum is electron scattering (which is much easier to model). Using their observations, they determined three times of the primary minimum and, combining them with times of the primary minimum from \cite{sem68}, came to the conclusion that clear indications of an orbital period change are absent. An important part of \cite{cher_khal73} was that the authors discovered significant physical variability during a night and from night to night.

\cite{khal73} published light curves of \mbox{V444~Cyg} in the $U$, $B$, and $V$ bands. He showed that these curves also exhibited significant physical variability, such that the shape of eclipses recorded on different nights changed. In this regard, the author criticized the work of Krohn and Gordon, who, when constructing average light curves, ``introduced corrections to the phases of observations on some nights, believing that the differences in individual curves obtained at different epochs were caused by random deviations in the ``center of gravity'' of the WR component's photometric disk''. Note that it was precisely these average light curves that \cite{sem68} used to determine the times of minima, which may thus contain systematic errors. By comparing broadband and narrow-band observations, Khaliullin showed that the times of minima determined from both are identical. He also pointed out that, in the presence of physical variability, the conventional method of constructing a mean light curve by averaging individual (formally equally accurate) measurements does not adequately account for physical variability and can lead to systematic errors in the phases of the mean light curve points. As an alternative, Khaliullin proposed the so-called ``group averaging'' method.

\cite{khal74} published a paper in which he first reported the detection of the orbital period change in \mbox{V444~Cyg}. His published table included times of minima from all the previous papers mentioned above. Considering their shortcomings, Khaliullin reprocessed the original data of all previous authors, using ``group averaging'' to obtain mean light curves and the Hertzsprung method to determine their phase shifts relative to each other. The table in \cite{khal74} included 11 values. In the same paper, the author presented theoretical formulas for the mass-loss rate of the WR component for various scenarios: conservative mass exchange with conservation of the system's total angular momentum, mass loss by the system through a spherically symmetric radial wind of the WR star, etc. Based on the calculated orbital period increase rate for each scenario, the article presented the corresponding mass-loss rates for the WR star. It was also shown that the hypothesis of period change due to apsidal shift in a system with non-zero eccentricity is inconsistent with observational data.

\cite{korn_cher79}, \cite{korn83} published new photoelectric observations of the system obtained in 1978 in narrow-band filters including only the continuum emission (\mbox{$\lambda~3100$}, 3520, \mbox{3780~\AA}). Using the methods of averaging individual measurements and determining the shifts of the average light curves proposed by \cite{khal73}, they added a new time of the primary minimum. In addition, at their request, Kurochkin analyzed old photographic observations of \mbox{V444~Cyg} using the SAI MSU photographic plates library. The mean epoch of these observations dates back to 1902, which significantly expanded the overall range of epochs. Using these data, the authors confirmed the results of \cite{khal74} and refined the rate of period change, as well as the mass-loss rate of the WR star in the scenario of mass loss by the WR stellar wind. They also showed that the ``third body'' hypothesis as the cause of the orbital period variations is inconsistent with the observational data. Beginning with this work, mass loss by the WR stellar wind is considered the primary cause of the orbital period increase of \mbox{V444~Cyg}.

\cite{eaton_cher_khal82} published two new times of minima (one for the primary and one for the secondary) based on observations with the {\it IUE} satellite. They used light curves obtained with the Fine-Error-Sensor (FES: a broadband detector, \mbox{$\lambda_{\rm eff}\sim 5500$~\AA}) and a light curve extracted from ultraviolet spectra in the range from ~1565 to \mbox{1900~\AA}, composed of four bands chosen to avoid the presence of spectral lines. The mean wavelength was \mbox{$\lambda\sim 1740$~\AA}. The authors note that these times are ``not very reliable'' due to the low accuracy of the measurements.

\cite{khal84} published another series of photoelectric observations of the system obtained in 1983 in the $V$ band and determined the corresponding time of the primary minimum. Using 14 times of minima (the times were listed in their previous work, see above; the times from \cite{eaton_cher_khal82} were not used), the authors determined a new, refined value for the rate of period change \mbox{$\dot{P} = (0.202\pm 0.018)$~s/year}.

\cite{under90} redefined the rate of period change of \mbox{V444~Cyg} using 16~photometric times of the primary minimum and one found from the radial velocity curve. Of the 16~photometric times of minima, nine were taken from \cite{sem68}. Three were taken from \cite{cher_khal73}, and one from \cite{khal84}. One time of the primary minimum was found from observations of \mbox{1964-66} by \cite{kuhi68} who used a narrow-band photoelectric spectral scanner with \mbox{$\lambda_{\rm eff}=4786$~\AA}. Finally, two more times were obtained by the authors themselves from photoelectric observations of 1988 in the $V$ band. The resulting value of the rate of period change was \mbox{$\dot{P} = (0.0842\pm 0.0011)$~s/year}. This value is significantly lower than the value obtained by \cite{khal84}. The reason for this is likely that, firstly, the authors did not use the Gaposhkin's and Kurochkin's times of minima, which significantly expand the range of observed epochs, and secondly, more than half of the times of minima \mbox{(9 out of 16)} in their paper were taken without any modifications from \cite{sem68}. As \cite{khal73} showed, these times may be distorted due to unaccounted systematic effects and incorrect assumptions by Kron and Gordon (whose data Semeniuk used) regarding the causes of the phase shift of individual points in the average light curves.

\cite{eaton_henry94} performed $B$, $V$ photometry of the secondary minimum of the light curve over two nights and determined the corresponding time of minimum using the template light curve from~\cite{cher_khal73}. They also determined another time of minimum in the same way from observations of the ascending branch of the primary and descending branch of the secondary minimum of the light curve obtained with the {\it IUE} satellite using the FES detector by \cite{breger85}. In addition, they used the times of minima published by \cite{khal84}, the original times from \cite{under90}, two ``unreliable'' times from \cite{eaton_cher_khal82}, and one time of the secondary minimum determined from the polarization curve of \cite{louis93}. Using these times of minima, they determined the rate of period change to be \mbox{$\dot{P} = (0.121\pm 0.006)$~s/year}, a value between those of Khaliullin et al. and Underhill.

\cite{agerer01} published times of minima for a number of binaries, including one for \mbox{V444~Cyg}, obtained from $B$, $V$ observations in \mbox{1999$-$2000}. \cite{eris11} published a paper in which they presented a new time of the secondary minimum, obtained from $B$, $V$ observations in 2007. They used a variant of Hertzsprung's method from \cite{kwee56} and the ``descending branch of the secondary minimum and the beginning of the ascending branch of this minimum'' to determine the time. Note that \cite{kwee56} specifically emphasized that for the correct determination of the time of minimum it is important to carefully choose the phase interval, otherwise, if minima of the light curve are not completely symmetric, the results may be distorted by systematic errors. As we will see below, this is probably what happened in this case. The table of times of minima in \cite[Table 6]{eris11}, used by the authors to determine the new ephemeris formula, contains \mbox{38~values}. 13~of them are taken from \cite{khal84}, and 17~--- from \cite{under90}. Thus, the data from these two papers account for \mbox{30~values} out of 38. As follows from our review above, the overwhelming majority of the values in Underhill's paper are based on the same original light curves as the values from Khaliullin et al. Underhill took these data from Semeniuk's paper and from works by Khaliullin, Kornilov, and Cherepashchuk of different years. Semeniuk used the light curves of Kron and Gordon, Hiltner, and others. Khaliullin and colleagues used the same original light curves, but processed the light curves from the earlier works of Kron and Gordon and others independently. For this reason, the times of minima in~\cite{khal84} and \cite{under90} do not formally coincide. The reason is that, when analyzing a given light curve obtained in a certain date range, different authors selected different orbital cycles for representative times of minima. Obviously, such times cannot be considered independent. Such formally different times, but based on the same original data, make up a significant portion of \mbox{Table~6} in~\cite{eris11}. This, strictly speaking, makes the authors' estimates of the parameters of the quadratic ephemeris formula and the estimate of the mass-loss rate of the WR~star invalid (although the obtained numerical values do not differ significantly from the ones obtained by other authors simply because the times of minima obtained by different authors from the same data are more or less consistent with each other).

\cite{jan16} published the light curve of \mbox{V444~Cyg} and a new time of its primary minimum, obtained from $V$-band observations carried out in 1990. The authors do not report the method by which they determined the time of the primary minimum, but their Fig.~1 shows that the minimum was approximated by a parabola. This method is guaranteed to introduce a systematic error, since the shape of the primary minimum differs from a parabola and is somewhat asymmetric. As will be shown below, such a (significant) systematic error in the time obtained by the authors is indeed present.

\cite{laur17} published a review of the variability of stars in OB associations in the $B$ band. Observations were made in \mbox{2011$-$2013}. The light curve of \mbox{V444~Cyg}, among others, is available online (the authors did not determine the times of minima).

\cite{shap23} published the latest work on this topic, in which the authors presented a new time of the primary minimum based on their observations in the $B$ band obtained in \mbox{2022$-$23}. They also drew on some data that had escaped the attention of previous researchers. Among them are the $B$-band light curve obtained by \cite{mof86} in 1984 and the light curve obtained by \cite{march98} with the HIPPARCOS satellite between 1989 and 1993 in a broadband filter (\mbox{$\lambda_{\rm eff}=5200$~\AA}, \mbox{$\rm FWHM\sim 2200$~\AA}). A distinctive feature of this work is that the authors did not use previously published times of minima but re-determined them from the original light curves using a modified Hertzsprung method. In particular, they did this for the data from \cite{eris11}, \cite{jan16}, \cite{mof86}, \cite{march98}, \cite{laur17}. This is what made it possible to identify systematic errors in the values of the times of minima given by \cite{eris11} and \cite{jan16}. This approach, on the one hand, made it possible to obtain the times of minima in a maximally uniform manner, but on the other hand, it left behind those times of minima given by authors who did not publish their original light curves. As a result, the number of photometric times of minima in~\cite{shap23} was~16, and the found rate of period change \mbox{$\dot{P} = (0.119\pm 0.003)$~s/year}.

\section{Updated ephemeris formula and mass-loss rate of the Wolf-Rayet star}\label{sec:ephem}

\mbox{Table~\ref{tab:moments}} lists the times of primary minimum for the light curve of \mbox{V444~Cyg} that we used in this paper to determine the updated ephemeris formula. Most of the values in the table were selected from the work of previous authors based on the following criteria: \mbox{(1)~the} list of times should be as complete as possible; \mbox{(2)~all} values should be obtained from independent original light curves; \mbox{(3)~they} should be obtained in as uniform a manner as possible. This task required the detailed study and review of previous work given above. We made every effort to find all previous publications relevant to this topic. Since many of the original light curves have been repeatedly reprocessed by different authors, in the fourth and fifth columns of the table we provide references to the papers from which we took a given value, as well as references to the papers where the original light curves (or times of minima) were published. Most of the times of minima related to the period after 1990 are taken from \cite{shap23}. Unfortunately, the authors of the paper do not present the values of times of minima themselves, but only the deviations \mbox{(O$-$C)} of the observed times from the times calculated using the linear formula (for a constant period). Calculating the times of minima requires this linear formula, which is not given in the paper. Fortunately, the authors of~\cite{shap23} provided us with this formula. Note that almost all the times of minima selected according to our criteria were determined using the Hertzsprung method, which ensures the homogeneity of the methodology. The only exception is one value of time from \cite{agerer01}, where the authors did not indicate the method used, and the light curves are not publicly available. \cite{eaton_henry94} and \cite{eris11} observed the system at the secondary minimum and cited the times of this minimum in their papers. For convenience, in \mbox{Table~\ref{tab:moments}}, such values are recalculated into times of the primary minimum.

%Table 1

\begin{table*}
\scriptsize
%\setcaptionmargin{0mm}
%\onelinecaptionsfalse
%\captionstyle{flushleft}

\captionsetup{margin=0mm}
\captionsetup{singlelinecheck=off}
\captionsetup{justification=raggedright}

\caption{Times of the primary minimum of the light curve of \mbox{V444~Cyg}}
\bigskip
\begin{tabular}{r|c|c|r|r}
\hline
 N   &      HJD        & $\sigma$   & Ref~1                   &  Ref~2  \\
 \hline
 1  ~&~  2415902.585  ~&~  0.0700  ~&  \cite{khal84}         ~&  (a)  \\
 2  ~&~  2423000.421  ~&~  0.0500  ~&  \cite{khal84}         ~&  \cite{gaposh41}  \\
 3  ~&~  2429900.294  ~&~  0.0040  ~&  \cite{khal84}         ~&  \cite{kron43}  \\
 4  ~&~  2430700.643  ~&~  0.0050  ~&  \cite{khal84}         ~&  \cite{kron43,kron50}  \\
 5  ~&~  2432103.376  ~&~  0.0040  ~&  \cite{khal84}         ~&  \cite{kron50}  \\
 6  ~&~  2432701.537  ~&~  0.0060  ~&  \cite{khal84}         ~&  \cite{hiltner49}  \\
 7  ~&~  2437903.892  ~&~  0.0030  ~&  \cite{khal84}         ~&  \cite{cher65}   \\
 8  ~&~  2438603.167  ~&~  0.0060  ~&  \cite{khal84}         ~&  \cite{gusein65}  \\
 9  ~&~  2438636.887  ~&~  0.0040  ~&  \cite{under90}        ~&  \cite{kuhi68}  \\
10  ~&~  2439003.341  ~&~  0.0040  ~&  \cite{khal84}         ~&  \cite{cher69}   \\
11  ~&~  2439003.346  ~&~  0.0040  ~&  \cite{khal84}         ~&  \cite{cher69}   \\
12  ~&~  2440903.163  ~&~  0.0030  ~&  \cite{khal84}         ~&  \cite{cher_khal72}  \\
13  ~&~  2441164.342  ~&~  0.0030  ~&  \cite{khal84}         ~&  \cite{khal73}  \\
14  ~&~  2443700.254  ~&~  0.0050  ~&  \cite{khal84}         ~&  \cite{korn_cher79}  \\
15  ~&~  2444913.420  ~&~  0.0040  ~&  \cite{eaton_henry94}  ~&  \cite{eaton_cher_khal82}   \\
16  ~&~  2444913.424  ~&~  0.0040  ~&  \cite{eaton_henry94}  ~&  \cite{eaton_cher_khal82}   \\
17  ~&~  2445528.460  ~&~  0.0030  ~&  \cite{khal84}         ~&  \cite{khal84}   \\
18  ~&~  2445890.740  ~&~  0.0500  ~&  \cite{shap23}         ~&  \cite{mof86}  \\
19  ~&~  2445970.754  ~&~  0.0060  ~&  \cite{eaton_henry94}  ~&  \cite{eaton_henry94}   \\
20  ~&~  2447394.562  ~&~  0.0060  ~&  \cite{under90}        ~&  \cite{under90}   \\
21  ~&~  2447773.678  ~&~  0.0060  ~&  \cite{under90}        ~&  \cite{under90}   \\
22  ~&~  2448182.315  ~&~  0.0040  ~&  \cite{shap23}         ~&  \cite{jan16}  \\
23  ~&~  2448355.026  ~&~  0.0170  ~&  \cite{shap23}         ~&  \cite{march98}  \\
24  ~&~  2448818.398  ~&~  0.0200  ~&  (b)                   ~&  (b)  \\
25  ~&~  2449256.491  ~&~  0.0200  ~&  (b)                   ~&  (b)  \\
26  ~&~  2449517.674  ~&~  0.0100  ~&  \cite{eaton_henry94}  ~&  \cite{eaton_henry94}   \\
27  ~&~  2449538.724  ~&~  0.0020  ~&  \cite{eaton_henry94}  ~&  \cite{eaton_henry94}   \\
28  ~&~  2451059.427  ~&~  0.0020  ~&  (b)                   ~&  (b)  \\
29  ~&~  2451413.275  ~&~  0.0080  ~&  \cite{agerer01}       ~&  \cite{agerer01}   \\
30  ~&~  2454328.347  ~&~  0.0080  ~&  \cite{shap23}         ~&  \cite{eris11}  \\
31  ~&~  2456362.995  ~&~  0.0020  ~&  \cite{shap23}         ~&  \cite{laur17}  \\
32  ~&~  2460019.452  ~&~  0.0170  ~&  \cite{shap23}         ~&  \cite{shap23}   \\
 \hline
\end{tabular}

\flushleft\scriptsize {\bf Note.} The table includes heliocentric dates and their uncertainties, as well as references to the articles from which the times of minima were taken (Ref~1) and to the articles containing the original data by which they were determined (Ref~2). (a)~--- Data of \mbox{N.~E.~Kurochkin}, private communication. (b)~--- This work, see text.

\label{tab:moments}
\end{table*}

%Fig.1

\begin{figure*}
\captionsetup{margin=0mm}
\captionsetup{singlelinecheck=off}
\captionsetup{justification=raggedright}
\includegraphics[width=0.9\textwidth]{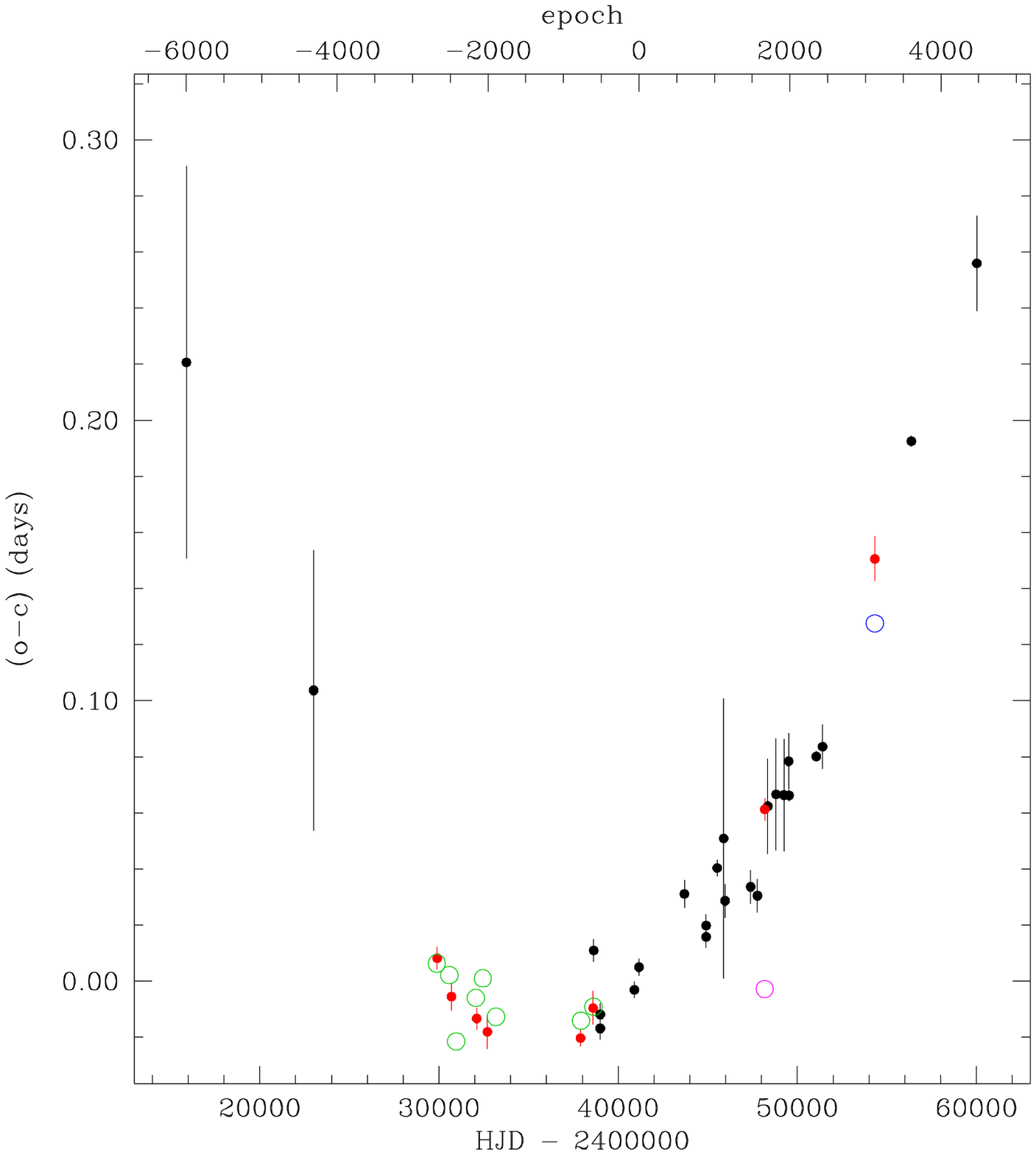}
\caption{\mbox{($O-C$)} values obtained for the times of primary minimum from \mbox{Table~\ref{tab:moments}} (dots with error bars), as well as \mbox{($O-C$)} values computed directly from the data of Semeniuk~\cite{sem68} (the date interval \mbox{$\sim$\,2430000$-$2439000}, shown by green circles), Janiashvili and Urushadze~\cite{jan16} (\mbox{$\sim$\,2448182}, the purple circle), Eris and Ekmeksi~\cite{eris11} (\mbox{$\sim$\,2454328}, the blue circle). The corresponding \mbox{($O-C$)} values based on times of primary minimum from \mbox{Table~\ref{tab:moments}}, and calculated using the same original data by Khaliullin et al.~\cite{khal84} and Shaposhnikov~\cite{shap23}, are shown in red. All \mbox{($O-C$)} values were calculated using the linear formula for the times of primary minimum given in the text.}
\label{fig:oc_comp}
\end{figure*}

In addition, the table includes three times of the primary minimum obtained by the author of this article by the Hertzsprung method from observations in 1992, 1993 and 1998 obtained with the \mbox{60-cm} telescope of the Crimean station of the SAI MSU in $B$, $V$, $R$ bands and in a narrow-band filter including the spectral region free of lines, with \mbox{$\lambda_{\rm eff} = 4270$~\AA}, \mbox{$\rm FWHM\sim 50$~\AA}.

Fig.~\ref{fig:oc_comp} shows the deviations \mbox{($O-C$)} of the observed heliocentric times of the primary minimum of the \mbox{V444~Cyg} light curve  from those calculated using the linear ephemeris formula from~\cite{khal84}:

$$
{\rm Min~I (HJD)} = 2441164.337 + 4.212435^d \cdot n\,,
$$

where $n$~is the number of orbital cycles elapsed since the initial epoch. In the review above, it was noted that \cite{khal73} criticized the methodology for determining the times of minima in earlier works, in particular, \cite{sem68}. We also noted that the methods for determining the times of minima in recent works \cite{eris11} and \cite{jan16} raise doubts. The figure shows a comparison of the times (actually the \mbox{(O$-$C)} values obtained with these times) given by \cite{sem68}, \cite{eris11}, \cite{jan16} with the times determined from the same data by \cite{khal84} and \cite{shap23}. The former are shown by open circles of different colors, the latter by red dots (their uncertainties are also shown). It is evident that the values of the times of minima determined by Khaliullin using ``group averaging'', designed to compensate for physical variability, have a smaller scatter than the Semeniuk values. However, it should be noted that at the time these authors carried out their work, the overall observation interval was comparatively short. In this situation, achieving maximum accuracy in the times of minima estimates was critically important. In the half-century that has passed since then, a significant amount of additional data has been accumulated. A present time, the poorer accuracy of the Semeniuk's values is no longer so critical from the standpoint of the fundamental conclusion that the orbital period of the system is increasing. Of course, \mbox{Table~\ref{tab:moments}} contains the times found by \cite{khal84}. Regarding the values of times from recent papers~\cite{eris11} and \cite{jan16}, it is clear from the figure that the values given by their authors fall outside the general behavior of the observed deviations \mbox{($O-C$)}. This confirms the validity of the doubts expressed regarding these values in our review. \mbox{Table~\ref{tab:moments}} presents the values of times of minima calculated from the original light curves of \cite{eris11}, \cite{jan16} by \cite{shap23}.

\begin{figure*}
\captionsetup{margin=0mm}
\captionsetup{singlelinecheck=off}
\captionsetup{justification=raggedright}
\includegraphics[width=0.9\textwidth]{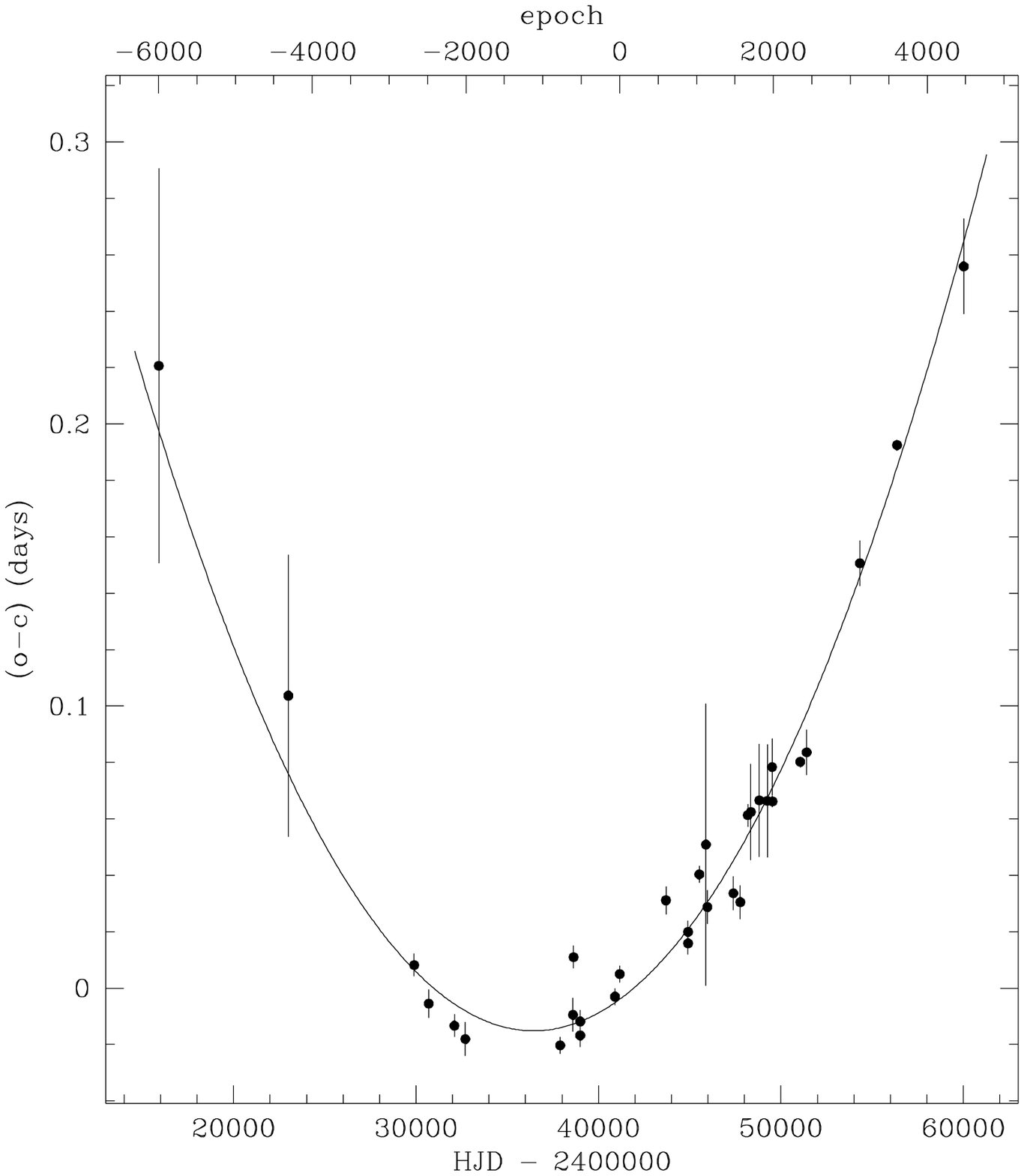}
\caption{Approximation of the \mbox{($O-C$)} values by a quadratic function. Only the \mbox{($O-C$)} calculated from the values from \mbox{Table~\ref{tab:moments}} were used.}
 \label{fig:oc_fit}
\end{figure*}

Fig.~\ref{fig:oc_fit} shows the approximation of the \mbox{($O-C$)} values calculated from the data in \mbox{Table~\ref{tab:moments}} by a quadratic function. The approximation parameters are given in \mbox{Table~\ref{tab:pars}}.

%Fig.2

The rate of period change can be calculated using the formula \mbox{$\dot{P} = 2A/P$}, for the value of~$A$ which we found, \mbox{$\dot{P} = 0.134\pm 0.003$~s/year}\footnote{Note that in \cite{shap23} \mbox{($O-C$)} values were computed in fractions of the period, therefore, to compare our parameter~$A$ and the parameter~$A$ from~\cite{shap23}, the latter must be multiplied by $P$.}. This value is in the range between the value of \cite{khal84} (\mbox{$\dot{P} = 0.202\pm 0.018$~s/year}) and the value of \cite{shap23} (\mbox{$\dot{P} = 0.119 \pm 0.003$~s/year}).

In the scenario of a change in the orbital period due to the loss of matter through the spherically symmetric wind of the WR star, the mass-loss rate by this star can be estimated using a simple formula~\citep{khal74}:

$$
\dot{M}_{\rm WR} = \frac{\dot{P}}{2P}(M_{\rm WR} + M_{\rm O} )\,,
$$

where $M_{\rm WR}$ and $M_{\rm O}$ are the masses of the WR and O stars, respectively. This formula assumes that the radii of the system's components are small compared to the size of the orbit, which is a reasonable approximation for \mbox{V444~Cyg} \mbox{($R_{\rm WR}\sim 4\,R_\odot$}, \mbox{$R_{\rm O}\sim 8.5\,R_\odot$}, \mbox{$a\sim\!36\,R_\odot$} (\citealp{ant01}; \citealp{shap23}). The values of $M\sin^3 i$ for both components of the system are determined from the radial velocity curves. The most recent spectroscopic study on this topic was published in~\cite{shap23}. The authors give the following estimates: \mbox{$M_{\rm WR}\sin^3 i = (10.0\pm 0.9)\,M_\odot$}, \mbox{$M_{\rm O}\sin^3 i = (24.7\pm 0.3)\,M_\odot$}. An analysis of the light curves of \mbox{V444~Cyg}, performed by different authors (see, e.g.,~\citealp{cher66, ant97, ant01}), gives an estimate of the orbital inclination: \mbox{$i = 78^\circ \pm 1.5^\circ$}. Substituting these values into the formula given above, we obtain the estimate \mbox{$\dot{M}_{\rm WR} = (6.82\pm 0.26) \times 10^{-6}\,M_\odot$/year}. This value is somewhat smaller than the value obtained by \cite{khal84}, \mbox{$\dot{M}_{\rm WR} = (1.02\pm 0.20) \times 10^{\rm -5}\,M_\odot$/year}. Our estimate, based on carefully filtered and most complete data on the period variation of \mbox{V444~Cyg}, can be considered the most reliable dynamical estimate of the mass-loss rate of the WR star to date.

%Table 2

\begin{table}
%\setcaptionmargin{0mm}
%\onelinecaptionsfalse
%\captionstyle{flushleft}

\captionsetup{margin=0mm}
\captionsetup{singlelinecheck=off}
\captionsetup{justification=raggedright}

\caption{Parameters of the quadratic formula for the times of primary minimum of the \mbox{V444~Cyg} light curve}
\bigskip
\begin{tabular}{rcl}
\hline
Min~I~(HJD) & $=$  & $E_0 + P \cdot n + A \cdot n^2$ \\
$E_0$       & $=$ & $2441164.333 \pm 0.001$, days \\
$P$         & $=$ & $4.2124550 \pm 0.0000005$, days \\
$A$         & $=$ & $(8.94 \pm 0.21) \times 10^{-9}$, days\\
 \hline
\end{tabular}
\label{tab:pars}
\end{table}

\section{Conclusions}\label{sec:concl}

In this paper, we carefully analyzed all currently available data on the period variations of \mbox{V444~Cyg}, based on the photometric light curves of the system. Particular attention was paid to ensuring that the values of the times of primary minimum found by different authors were based on independent original light curves. Among the authors whose results were based on the same original light curves, we selected those who analyzed these data in the most uniform manner possible. By adding three new independent estimates obtained by us, to the known times of primary minimum, we compiled a final table, including 32 values covering the maximum currently available interval of observations of the system. Using this table, we found updated parameters of the quadratic formula for the times of primary minimum of the system. We believe that this formula is currently the most reliable in the entire history of studies of the period variations of \mbox{V444~Cyg}. We hope that it will be useful in future studies of the system.

The found value of the rate of period increase made it possible to obtain a new estimate of the mass-loss rate by the WR star in the scenario of system mass loss through a spherically symmetric WR wind.
\medskip

\section*{FUNDING}

The author expresses gratitude for the support of this work to the Russian Science Foundation (project 23-12-00092).

\section*{CONFLICT OF INTEREST}

The author of this work declares that he has no conflicts of interest.

\bibliographystyle{aspb3}
\bibliography{Antokhin}

\end{document}